\documentclass{article}
\usepackage{arxiv}

\usepackage[utf8]{inputenc} 
\usepackage[T1]{fontenc}    
\usepackage{hyperref}       
\usepackage{url}            
\usepackage{booktabs}       
\usepackage{amsfonts}       
\usepackage{nicefrac}       
\usepackage{microtype}      
\usepackage{lipsum}		
\usepackage{graphicx}
\usepackage{natbib}
\usepackage{doi}
\usepackage{amsmath}
\usepackage{amssymb}
\usepackage{mathtools}
\usepackage{amsthm}
\usepackage{threeparttable}
\usepackage{comment}
\usepackage{colortbl}
\usepackage{xcolor}
\usepackage{multirow}
\usepackage{listings}
\usepackage{color}
\definecolor{HL}{RGB}{255,0,255}
\definecolor{Blue}{RGB}{0,0,255}

\title{Decomposing weather forecasting into advection and convection with neural networks}

\author{Mengxuan Chen\\
	Tsinghua University\\
	\texttt{chenmx21@mails.tsinghua.edu.cn} \\
	\And
	Ziqi Yuan \\
	Tsinghua University\\
	\texttt{yzq21@mails.tsinghua.edu.cn} \\
	\AND
	Jinxiao Zhang \\
	Tsinghua University \\
	\texttt{zhang-jx22@mails.tsinghua.edu.cn} \\
	\And
	Runmin Dong \\
	Tsinghua University\\
	\texttt{drm@mail.tsinghua.edu.cn} \\
	\And
	Haohuan Fu\thanks{Correspondence to: H. Fu, haohuan@tsinghua.edu.cn}\\
	Tsinghua University\\
	\texttt{haohuan@tsinghua.edu.cn}\\
}

\date{}

\renewcommand{\shorttitle}{}

\begin{document}
\maketitle

\begin{abstract}
Operational weather forecasting models have advanced for decades on both the explicit numerical solvers and the empirical physical parameterization schemes. However, the involved high computational costs and uncertainties in these existing schemes are requiring potential improvements through alternative machine learning methods. Previous works use a unified model to learn the dynamics and physics of the atmospheric model. Contrarily, we propose a simple yet effective machine learning model that learns the horizontal movement in the dynamical core and vertical movement in the physical parameterization separately. By replacing the advection with a graph attention network and the convection with a multi-layer perceptron, our model provides a new and efficient perspective to simulate the transition of variables in atmospheric models. We also assess the model's performance over a 5-day iterative forecasting. Under the same input variables and training methods, our model outperforms existing data-driven methods with a significantly-reduced number of parameters with a resolution of 5.625$^{\circ}$. Overall, this work aims to contribute to the ongoing efforts that leverage machine learning techniques for improving both the accuracy and efficiency of global weather forecasting.
\end{abstract}

\section{Introduction}\label{introduction}
Forecasting weather accurately is vital for human society. Numerical Weather Prediction (NWP), a dominant approach for weather forecasting, has been developed for more than 100 years \cite{bauer2015quiet}. Benefited enormously from the advances in supercomputing capacity and observations, this revolution spans from parameterization schemes representing unresolved physical processes to data assimilation providing more precise initial states, from ensemble modeling addressing minor perturbations in initial conditions to increasing resolutions for eliminating uncertainties. Operational systems such as the Integrated Forecasting System (IFS) from the European Centre for Medium-Range Weather Forecasts (ECMWF) and the Global Forecast System (GFS) from the National Centers for Environmental Prediction (NCEP), play a crucial role in decision-making and risk management in everyday life.

\begin{figure}[h!]
\begin{center}
    \centerline{\includegraphics[width=0.5\columnwidth]{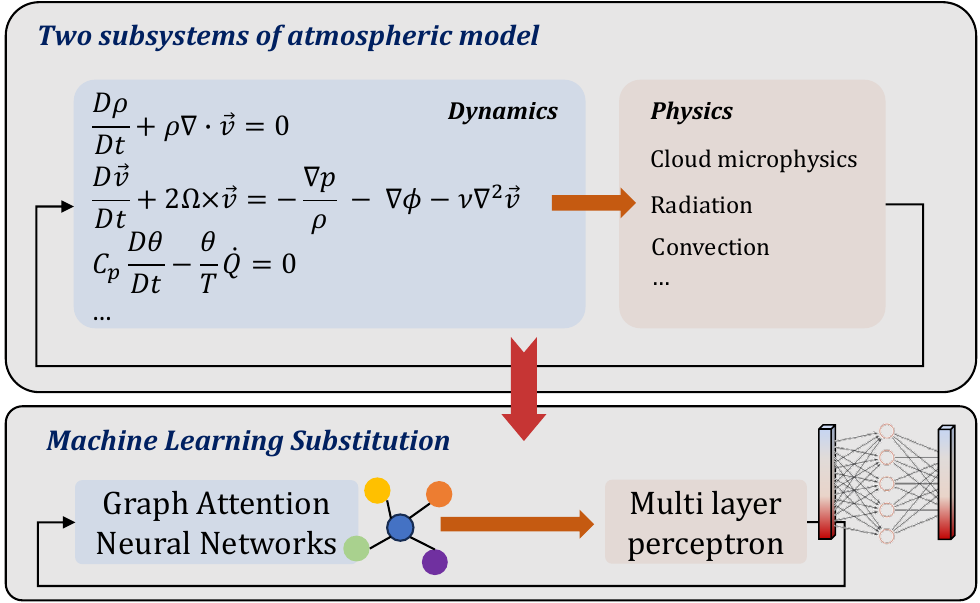}}
    \caption{The combination of dynamics and physics forms the comprehensive atmosphere model. This study proposes a modularized weather forecasting model combining Graph Attention Network and Multi-Layer Perceptron (GAT-MLP) to simulate the advection and convection in the atmospheric model respectively, in an iterative forecasting manner.}
    \label{fig:task}
    \vskip -0.2in
    \end{center}
\end{figure}

A typical atmospheric model, such as the Community Atmosphere Model (CAM) or the Weather Research \& Forecasting Model (WRF), typically comprises two parts: dynamical core and physical parameterization schemes. The Navier-Stokes equations, the mass continuity equations, and the governing equations of thermodynamics and ideal gas, form the foundation of numerical models in the dynamical core. In addition to these explicit equations, sub-grid or complex physical processes that cannot be resolved by the numerical models are calculated in the empirical-based physical parameterization schemes, such as the radiation transport and convection formation, as depicted in Figure \ref{fig:task}. The combination of dynamics and physics forms the comprehensive atmospheric model. In pursuing a higher resolution and more precise description of the atmospheric state evolution, NWP faces challenges in terms of computational resources and the complex nonlinear system \cite{randall2013beyond,kriegmair2022using}. Consequently, these challenges promote the development of machine learning (ML), as a substitution method for traditional weather forecasting models.

In these years, we witnessed the great success of using ML in topics related to weather forecasting, including but not limited to the data assimilation, substitution of climate models, and the correction of the model output \cite{buizza2022data, bonavita2020machine, watt2021correcting}. Starting from using Multi-Layer Perceptron (MLP) to simulate the complex nonlinear physical processes within each vertical column (grid with different pressure levels), ML shows great potential in learning intricate relationships from huge amount of data. To address the limitation that MLP cannot explicitly learn the spatial relationship between grids, Convolutional Neural Networks (CNN) are then introduced and widely used in weather forecasting to achieve better performance and higher computational efficiency. However, the characteristic of CNN viewing the atmospheric status as an image with multiple channels, is not in line with the real case. For instance, transferring information from the leftmost side to the rightmost side in an image poses challenges, and is only performed in the deep layers in CNN-based models, despite the expectation of continuity in the real world. Furthermore, there is no specific guarantee of the boundaries in the image, such as the north and south poles. Also, the convolutional kernels assume the translation invariance in the image, which is not always correct in reality. For example, the increase in sea surface temperature (SST) may have different impacts on its east and west coast \cite{palmer1984response}. To address the issues mentioned above, attention mechanisms, Transformer or Graph Neural Networks (GNN) are introduced in the weather forecasting task, to learn a better relationship between variables exchange at different locations. Particularly, GNN can explicitly model the interactions and variable transfer between the nodes by designing specific graph structures, which is more interpretable \cite{liu2022interpretability}, and has gained attention in recent years \cite{keisler2022forecasting, lam2023learning}.

Among all the successful studies using ML methods for weather forecasting, most try to use one unified neural network to represent both dynamics and physical processes in the atmospheric model. These models are meticulously designed to extract useful information from vast amounts of training data. Consequently, they experience exponentially-increased parameters when pursuing more accurate results. Simultaneously, the lack of interpretability of these complex models make it difficult to express the working mechanisms. Since atmospheric models have a long development history, it is another choice to draw inspiration and learn from the accumulated experience and insights instead of completely discarding the original methods when utilizing ML for substitution.

\begin{figure}[h!]
  \vskip 0.2in
  \begin{center}
  \centerline{\includegraphics[width=\columnwidth]{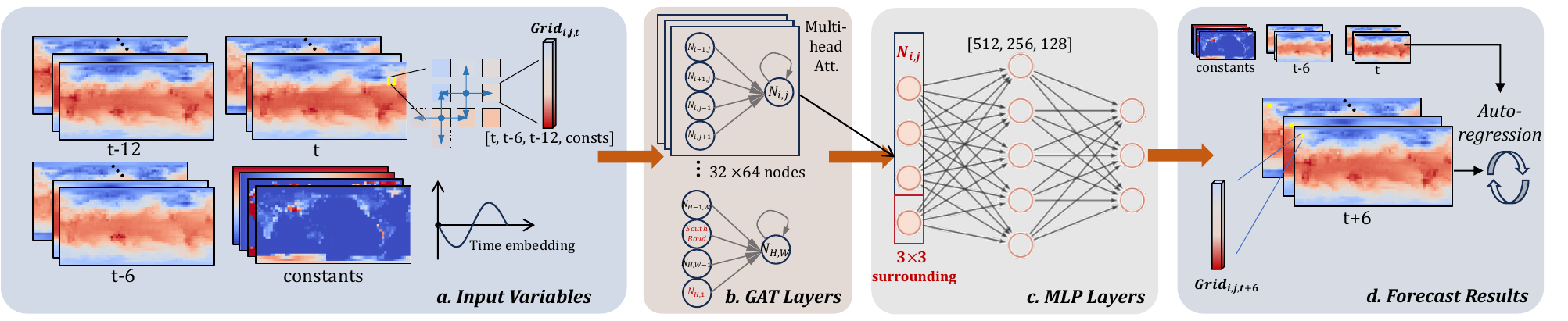}}
  \caption{A general view of our proposed model. Variables from the previous three timesteps and the constants are input variables. Each grid is viewed as a node and connected with four neighbors. Then, the whole graph is put into a 2-layer Graph Attention Network (GAT), followed by a 4-layer Multi-Layer Perceptron (MLP) to predict the variables in the next timestep. This model adopts the auto-regressive manner for prediction, which means, the predicted variables are fed back into the model for iterative forecasting.}
  \label{fig:model}
  \end{center}
  \vskip -0.2in
\end{figure}

Therefore, in this work, as shown in Figure \ref{fig:task}, we propose a simple yet effective model to learn and represent the dynamics and physics of the atmospheric model as two separate neural networks. Figure \ref{fig:model} shows a general workflow of our proposed model. Compared with previous works, our model provides the following unique features: 
\begin{itemize}
    \item \textbf{A divide-and-conquer modeling approach.} Different from using a comprehensive model to unify all the processes in the atmospheric model, our model uses two components to learn the advection in dynamical core and convection in physical parameterization part separately. We adopt a Graph Attention Network (GAT) \cite{velivckovic2017graph} to emulate the advection between the nodes and their connected neighbors and an MLP to simulate the physical processes within each vertical column.
    \item \textbf{Stencil-like graph structure.} Different from first mapping the grid data onto the icosahedron mesh, we directly model each latitude-longitude (lat-lon) grid as an independent node, and connect each node with its four neighbors through graph edges. Besides, two boundary nodes are set to define the north and south poles to ensure the closure of the graph structure.
    \item \textbf{Non-local physical process simulation}. Different from MLP models that perform in every vertical column, we integrate information from surrounding 3 $\times$ 3 grids to simulate the physical process with non-local MLP. 
\end{itemize}
Moreover, in contrast to existing models, our model is lightweight with only 4.38M parameters in total for a resolution of 5.625$^{\circ}$.

\section{Literature Review}\label{litreview}
\subsection{Machine Learning in Weather Forecasting}
Operational weather forecasting typically involves data preparation, data assimilation, prediction, postprocessing, and results evaluation. In order to replace the time-consuming computational process and better capture the non-linear relationship in atmospheric variables, ML-based weather forecasting models in current stage aim to replace the entire process except for data preparation and results evaluation \cite{schultz2021can}. 

The availability of extensive datasets like ERA5 \cite{ERA5} and model data from CMIP projections \cite{eyring2016overview} have significantly contributed to the growth of machine learning applications in weather forecasting and climate modeling, particularly in end-to-end workflows. At the very beginning, MLP and CNN are the most commonly used methods \cite{rasp2020weatherbench,weyn2020improving,rasp2021data,clare2021combining}. In recent years, more advanced methods have been gradually adopted in weather forecasting with the development of both computer vision techniques and hardware. In order to capture the temporal dependence, methods such as Recurrent Neural Networks (RNN) and Long Short-Term Memory (LSTM) are added to the network \cite{espeholt2022deep,chen2023swinrdm}. To model the long-range dependencies well, newly-developed models tend to use Transformer as the backbone to extract features \cite{pathak2022fourcastnet,bi2023accurate,nguyen2023climax,chen2023fengwu}. Besides, some studies have demonstrated the potential of utilizing Graph Neural Networks for large-scale and long-term weather forecasting \cite{keisler2022forecasting, lam2023learning} and using diffusion models for the ensemble prediction \cite{price2023gencast}.

\subsection{Graph Neural Networks in Earth Systems}
Graph Neural Networks (GNN) was initially proposed to deal with unstructured data, which needs to be represented in graph domains \cite{scarselli2008graph}. By learning node representations through iterative message-passing between interconnected nodes several times, GNN can capture complex dependencies and intricate interconnections within a graph. GNNs have gained growing attention over the past five years for their effectiveness in modeling the Earth system's dynamic mechanisms and variable exchanges. Starting with its application to enhance El Ni\~{n}o forecasting \cite{cachay2020graph}, GNNs have demonstrated their potential and aptness in modeling \textit{grid} climate data. This becomes particularly pertinent in scenarios where the current state in one grid is significantly influenced by the states from the neighbors.

After that, GNN and graph convolutional networks (GCN) are tested to simulate atmospheric radiative transfer in weather and climate models. The findings indicate that graph-based networks exhibit a smaller mean biased error and greater flexibility when compared to models constructed with CNN and MLP \cite{cachay2021climart}. In 2022, Keisler employs the GNN for global weather forecasting \cite{keisler2022forecasting}. The study demonstrates that message-passing GNN outperformed CNN-based weather forecasting models in terms of both accuracy and model size during a 6-day iterative running. Additionally, the GNN performance was comparable to that of the European Centre for Medium-Range Weather Forecasts (ECMWF) model. As a superior follow-up, the GraphCast model elevates the GNN-based medium-range global weather forecasting to a new level, and outperforms ECMWF's High RESolution forecast (HERS), one of the most accurate operational deterministic systems in the world \cite{lam2023learning}. Both of the above-mentioned works project the climate data onto the icosahedron meshes to eliminate the artifacts of discontinuity and no specific north and south poles in the original lat-lon grid data. 

Beyond these works, GNNs have been applied in many other fields, such as the heatwave prediction \cite{li2023regional}, river network learning \cite{sun2022graph}, hourly solar radiation prediction \cite{gao2022interpretable}, and PM$_{2.5}$ forecasting \cite{wang2020pm2}. The works mentioned above show that GNN has great potential in applications related to the Earth systems, especially in learning the relationships between different variables and locations.

\section{Approach}  
\subsection{Model Architecture}
The complete workflow for our weather forecasting model includes a grid-to-node encoder, a GAT-MLP model to simulate the advection and convection in the atmospheric models, and a node-to-grid forecast module, as depicted in Figure \ref{fig:model}. 

\paragraph{Grid-to-Node Encoder} The grid-to-node encoder is designed to formulate each grid with different pressure levels in the lat-lon coordinate as a node in the graph, and the neighborhood relationship between them is represented with graph edges, as shown in Figure \ref{fig:model}(a). We use data from previous three time steps as the major input variables. Different from previous GNN-based weather forecasting approaches  \cite{keisler2022forecasting,lam2023learning} that first project the original lat-lon grid ERA5 data onto icosahedron meshes with various resolutions, our model directly employs the lat-lon grid to eliminate the errors introduced by mesh transformation. The input variables are encoded as the node features.

We formulate the node's edges according to the 2D stencil computation. For a given node $N_{i,j}$ located in line $i$ and column $j$, it connects to its four neighbours $N_{i-1,j}$, $N_{i+1,j}$, $N_{i,j-1}$, $N_{i,j+1}$ to simulate the five-point stencil calculation. Besides, we connect nodes $\mathbf{N_{:,0}}$ to $\mathbf{N_{:,W-1}}$ to ensure continuity, align with the real world. The nodes $\mathbf{N_{0,:}}$ and $\mathbf{N_{H-1,:}}$ are linked to pseudo nodes representing the north and south poles, respectively, to ensure the closure. $H$ and $W$ denote the total number of grids in the latitude and longitude directions, which, in this case, are 32 and 64 for 5.625$^{\circ}$ resolution. The variables are encoded as node features, with the pseudo nodes having all-zero values. The node features are then fed into the GAT. Appendix \ref{grid2node_appendix} provides a detailed explanation of this process.

\paragraph{GAT-MLP Model Structure} In the atmospheric model, the dynamical core uses numerical methods to transform the atmospheric governing equations into discrete forms, and stencil calculation is used for calculating the fluxes in and out from neighboring grids. Some atmospheric variables that cannot be solved at the grid scale or be described by the atmospheric equations are represented by the physical parameterization schemes, which are calculated within each vertical column and do not exchange with neighboring grids. Inspired by the success of GNN in weather forecasting and of MLP in subgrid physical process representation, as well as the running mechanisms in atmospheric models, our model is primarily composed of two parts: a 2-layer GAT to simulate the advection between interconnected nodes, and a 4-layer MLP to mimic the convection within each vertical column, as illustrated in Figure \ref{fig:model} (b-c). 

\textbf{GAT Layer:} For a given node $N_i$ and one of its neighbour $N_j$, the attention coefficient first computes the similarity factor between itself and its neighbour with
\begin{equation}
    \alpha_{i,j} = \frac{\exp\left(\text{LeakyReLU} (f(
    W\cdot N_i || W\cdot N_j))\right)}{\sum_{j \in J}\exp(\text{LeakyReLU} (f(W\cdot N_i|| W\cdot N_j)))},
\end{equation}
where $W$ is the shared weighted matrix, $||$ denotes the concatenate process, and $J$ is the set of $N_i$'s neighbours, with a self-loop embedding. Then, the neighbour information is aggregated to $N_i$ to update its feature with
\begin{equation}
    N_i' = \sigma(\sum_{j \in J}\alpha_{i,j}W'N_j),
\end{equation}
where $\sigma$ is the activation function. In this work, we use the multi-head attention mechanism to improve the model performance by setting the number of heads to 8. Therefore, for each head, it gives the updated feature $N_i'$ with
\begin{equation}
    N_i'(k) = \sigma(\sum_{j \in J}\alpha_{i,j}^kW'^kN_j),
\end{equation}
for $k$ is the head index. We use a two-layer GAT to simulate the exchange process between interconnected nodes. Currently, the vertical dependence in dynamical part is not explicitly modeled, and we assume that the MLP layer will capture part of the dependence.

\textbf{MLP Layer}: A four-layer MLP with hidden dimensions of 512, 256 and 128 is used for the second stage, as previous studies showed that MLP with shallow layers is enough to simulate the physical process within vertical columns \cite{brenowitz2018prognostic,rasp2018deep,yuval2023neural,watt2024neural} Besides, \cite{wang2022non} has demonstrated that adding surrounding 3$\times$3 neighboring information help increase the performance. Therefore, for each node, the features from surrounding $3\times 3$ nodes are extracted with the one-layer CNN with the kernel size of $3\times 3$, and encoded into 256 hidden dimensions. Then, the extracted features are concatenated with the node features, and fed into the MLP network to simulate the physical process within each vertical column. The output of the MLP layer is the predicted variables at the next time step.

\paragraph{Node-to-Grid Forecast} The node-to-grid forecast module is shown in Figure \ref{fig:model}(d), which converts the graph structure to the original lat-lon coordinate. The same as the operational weather forecasting model, our proposed model follows the auto-regression forecast scheme. We use iterative forecasting with a step of 6 hours in the experiment. For example, $$z_{t+12} = \textit{Model}(\textit{Model}(z_{t}, z_{t-6}, z_{t-12}, \textit{consts}), z_{t}, z_{t-6}, \textit{consts}).$$ The forecast process is repeated until the desired lead time is reached. While iterative prediction may exhibit decreased performance compared to direct or continuous forecasts used in many studies attributed to the accumulation of prediction errors, it offers a more realistic approach to support long-term operations.

\subsection{Data Preprocessing}
\paragraph{Time and Location Embedding}\label{embed} As for the constant variable latitude and longitude in WeatherBench dataset, we embed the latitude and longitude as 
$$lat = \cos \frac{2\pi (h+90)}{180}, 
lon = \sin \frac{2\pi w}{360},$$
where $h \in [-90^{\circ}, 90^{\circ}]$ and $w \in [0^{\circ}, 360^{\circ}]$. Besides, the time is encoded as 
$$t' = -\sin \frac{2\pi}{T} t,$$
where $T$ is the total hours in each year, and $t$ is the specific hour in that year.

\paragraph{Data Normlization} We apply instance normalization \cite{ulyanov2017improved} to our input variables, using mean and standard deviation values calculated from the corresponding time steps for normalization. Variational mean and standard deviation values aim to enhance the model's robustness, particularly in adapting to scenarios like climate change, where temperature increases may introduce greater variability.

\subsection{Iterative Training} 
We train our model for 150 epochs, with an initial learning rate of 0.001 and the AdamW optimizer. The loss function used is the latitude-weighted RMSE, where the latitude-weight is represented as
\begin{equation}
  L(h) = \frac{\cos(h)}{1/H\sum_{lat=1}^{H}\cos(lat)}
  \label{eq:l-weighted}
\end{equation}
The learning rate is reduced by a factor of two after every ten epochs. During the initial 100 epochs, the model is trained to forecast 6 hours ahead (forecasting one step). Subsequently, epochs 101-125 are dedicated to forecasting 30 hours ahead (forecasting five steps), epochs 126-140 are used for 54 hours ahead (forecasting nine steps), and the final ten epochs are focused on forecasting 78 hours ahead (forecasting 13 steps). The multistep training process aims to improve the iterative forecasting performance.

\section{Experiments and Results}
\subsection{Data}
We employ WeatherBench1, a widely used preprocessed dataset for weather forecasting based on ECMWF's ERA5 dataset \cite{rasp2020weatherbench, ERA5}. We use the 5.625$^{\circ}$ resolution data from 1980 to 2018, with a sampling interval of 10 hours, and set the proportion of training, validation and test dataset to 8:1:1. Table \ref{tab:variables} lists the variables used in our model.

\vskip -0.15in 
\begin{table}[htbp]
  \centering
  \begin{threeparttable}[b]
  \caption{WeatherBench variables used in the model. Below the divider line are the constants.}
  \vskip 0.1in  
  \begin{tabular}{lll}
      \toprule
    Variable & Abbr. & Pressure levels \\
    \midrule
    geopotential & z     & Multi-levels\tnote{1} \\
    temperature & t     & Multi-levels \\
    u\_component\_of\_wind & u     & Multi-levels \\
    v\_component\_of\_wind & v     & Multi-levels \\
    vorticity & vo    & Multi-levels \\
    specific\_humidity & q     & Multi-levels \\
    2m\_temperature & t2m   & Single level \\
    total\_precipitation & tp    & Single level \\
    \midrule
    land-sea mask & lsm   & / \\
    Orography & h     & / \\
    latitude location & lat2d & / \\
    longitude location & lon2d & / \\
    \bottomrule
    \end{tabular}%
    \begin{tablenotes}
      \item[1] Multi-levels refer to 50, 250, 500, 600, 700, 850, 925 [hPa]
    \end{tablenotes}
    \label{tab:variables}%
  \end{threeparttable}
  \vskip -0.1in
\end{table}%

For our chosen subset of WeatherBench1, geopotential height can provide valuable information about the vertical structure, pressure distribution, and large-scale circulation of the atmosphere. $u$ and $v$ component of wind and vorticity are fundamental parameters in weather forecasting, providing critical information about the movement, convergence, and divergence of air masses, while 2m temperature is essential for human activity. Variables at time $t$, $t-6$, and $t-12$ are used as the input, and the task involved forecasting these variables at a lead time up to 120 hours iteratively, with an interval of 6 hours. Including variables from multiple time steps may help the model to capture the time dependencies and diurnal cycles, simultaneously having a more robust and stable performance in iterative forecasting tasks.

\subsection{Evaluation Metrics}
The evaluation metrics are latitude-weighted RMSE and Anomaly Correlation Coefficient (ACC), as described in Equation \ref{eq:rmse} and \ref{eq:acc}.
\begin{equation}
    RMSE = \frac{1}{N}\sum_{t=1}^{N}\sqrt{\frac{1}{HW}\sum_{h=1}^{H}\sum_{w=1}^{W}L(h)(y_{t,h,w}-\hat{y}_{t,h,w})^2}
    \label{eq:rmse}
\end{equation}
\begin{equation}
    ACC = \frac{\sum_{t,h,w}L(h)y'_{t,h,w}\hat{y'}_{t,h,w}}{\sqrt{\sum_{t,h,w}L(h)(y'_{t,h,w})^2}\sqrt{\sum_{t,h,w}L(h)(\hat{y'}_{t,h,w})^2}}
    \label{eq:acc}
\end{equation}
where $L(h)$ is the latitude-weight shown in Equation \ref{eq:l-weighted}. $y'$ and $\hat{y'}$ are respectively the anomalies of $y$ and $\hat{y}$ against the climatology $C_{h,w} = \frac{1}{N}\sum_{t=1}^{N}y_{t,h,w}$ at each grid.

\subsection{Global Forecasting Results}

\begin{figure}[h!]
  \vskip 0.1in
  \begin{center}
\centerline{\includegraphics[width=\columnwidth]{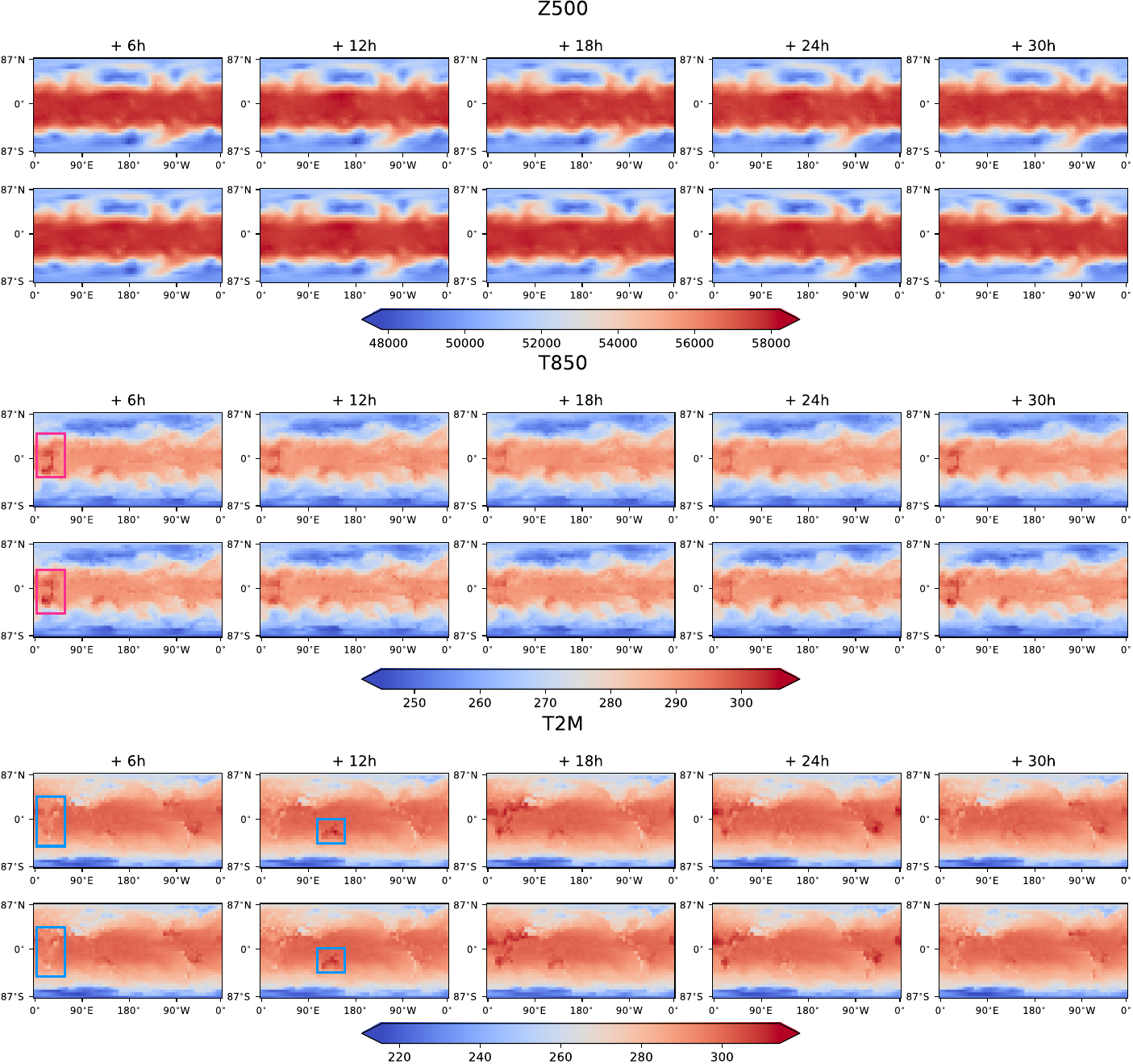}}
  \caption{An example of the visualization of global forecasting results with 30 hours lead time of Z500, T850, and T2M. For each variable, the first row shows the predicted results by our model, and the second row shows the ground truth from WeatherBench dataset.}
  \label{fig:forecasting_results}
  \end{center}
  \vskip -0.2in
\end{figure}

We assess the model performance in global weather forecasting. We report the metrics scores of geopotential at 500 hPa (Z500), temperature at 850 hPa (T850), and temperature near surface (T2M), and show the results of iterative forecasting. Figure \ref{fig:forecasting_results} provides a visual example of the global forecasting results. For each variable, the first row shows the prediction and the second row represents the ground truth from the WeatherBench dataset. While from the ERA5 data, Z500 does not exhibit significant variations in a short term, T850 and T2M show clear diurnal cycles as the forecast progresses, particularly noteworthy in regions such as Africa and Australia, highlighted by the pink and blue boxes in the figure. Our model successfully captures the intricate global atmospheric patterns well, and their changing cycles over time.

\subsection{Ablation Study}\label{ablation}
\begin{figure}[h!]
    \vskip 0.2in
    \begin{center}
    \centerline{\includegraphics[width=0.7\columnwidth]{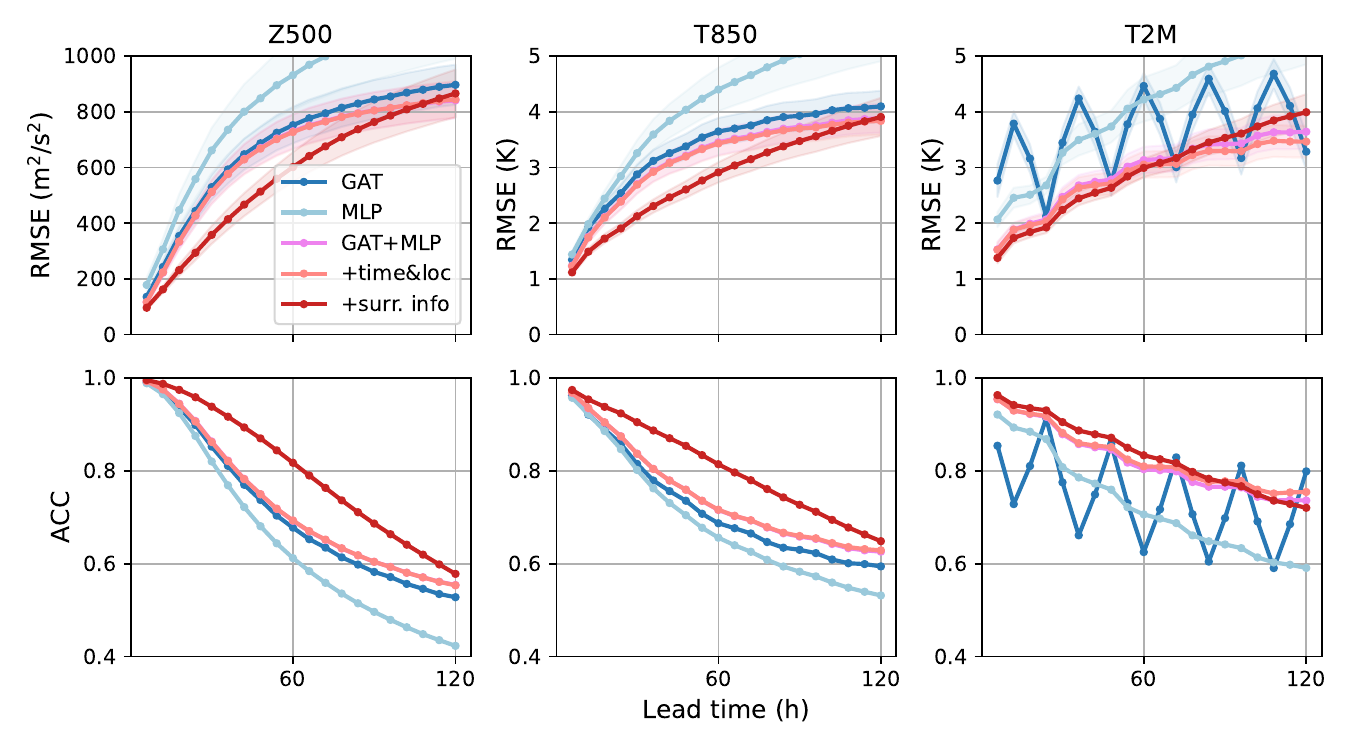}}
    \vskip -0.2in
    \caption{Ablation study results for 5-day iterative forecasting.}
    \label{fig:ablation}
    \end{center}
    \vskip -0.2in
\end{figure}

We assess the effectiveness of our model with the ablation study. The results for one-step prediction for Z500, T850, T2M and Total Precipitation (TP) are shown in Table \ref{tab:ablation}, and the whole iterative forecasting performance with a 5-day lead time is shown in Figure \ref{fig:ablation}. 

  \begin{table}[htbp]
    \centering
     \caption{Ablation study results for one-step prediction.}
     \vskip 0.1in
       \begin{tabular}{llllrrrrrrrr}
         \toprule
             &       &       &       & \multicolumn{2}{c}{Z500} & \multicolumn{2}{c}{T850} & \multicolumn{2}{c}{T2M} & \multicolumn{2}{c}{TP}   \\
       \cline{5-12}
       \multicolumn{4}{c}{} & \multicolumn{1}{c}{RMSE$\downarrow$} & \multicolumn{1}{l}{ACC$\uparrow$} & 
       \multicolumn{1}{l}{RMSE$\downarrow$} & \multicolumn{1}{l}{ACC$\uparrow$} & \multicolumn{1}{l}{RMSE$\downarrow$} & \multicolumn{1}{l}{ACC$\uparrow$} & \multicolumn{1}{l}{RMSE$\downarrow$} & \multicolumn{1}{l}{ACC$\uparrow$}  \\
       \midrule
       $\surd$ &       &       &       & 135.91 & 0.99  & 1.34  & 0.97  & 2.76  & 0.89  & 0.38  & 0.54  \\
      &     $\surd$  &       &       &  178.26 & 0.99  & 1.44 & 0.96  &  2.07 & 0.94 & 0.39 & 0.52 \\
       \midrule
       $\surd$ & $\surd$ &  &  & 116.01 &  0.99 &  1.24 & 0.97  & 1.53  &  0.97 &  0.37 & 0.56 \\  
       $\surd$ & $\surd$ & $\surd$ &  & 116.49 &  0.99 & 1.23&  0.97 & 1.52  &  0.97 &   0.37  &   0.56   \\
       $\surd$ & $\surd$ & $\surd$ & $\surd$ &  \textbf{96.30} & 1.00  & \textbf{1.12} & 0.98  &   \textbf{1.38} &  0.97  & \textbf{0.36}  &  0.61 \\
       \midrule
       \rowcolor{pink!30} GAT   & MLP & + time\&loc. & + surr. info.&       &       &       &       &       &       &     &    \\
       \bottomrule
       \end{tabular}%
     \label{tab:ablation}%
     \vskip -0.1in
   \end{table}%

Two baselines are set up to test the primary performance that can be obtained by using either GAT or MLP. The first baseline is a 2-layer GAT with 8 heads and 128 hidden units, denoted as \textit{GAT}. The second one is a four-layer MLP with 512, 256, and 128 hidden units, denoted as \textit{MLP}. From the results, the GAT forecasting results in T2M vibrate with the iterations, which means it can not capture the diurnal variations. In addition, the results for single MLP become divergent after several iterations. Using either component to simulate the weather forecasting process is still challenging.

By contrast, the combined model can alleviate the above-mentioned problems, and enhance the forecasting performance. \textit{+time\&loc.} involves the time and location embeddings described in Section \ref{embed}. Adding the time and location embeddings does not have much impact on the performance of Z500 and T850 (as two lines describing \textit{GAT+MLP} and \textit{+time\&loc.} in Figure \ref{fig:ablation} are overlapped). However, it can improve the forecasting of T2M after several iterations. This may indicate that the time embeddings can improve the learning of the temperature diurnal cycle. \textit{+surr. info.} indicates the inclusion of surrounding information from neighboring $3\times 3$ grids. Adding the information from neighbors can improve the model's performance within the iterative prediction of 5 days. However, the performance decays as the iteration continuous. The possible reason is that the uncertainty in surrounding information hampers the model's performance in long-term forecasting.

\subsection{Validation of Iterative Training}
\begin{figure}[h!]
    \begin{center}
    \centerline{\includegraphics[width=0.5\columnwidth]{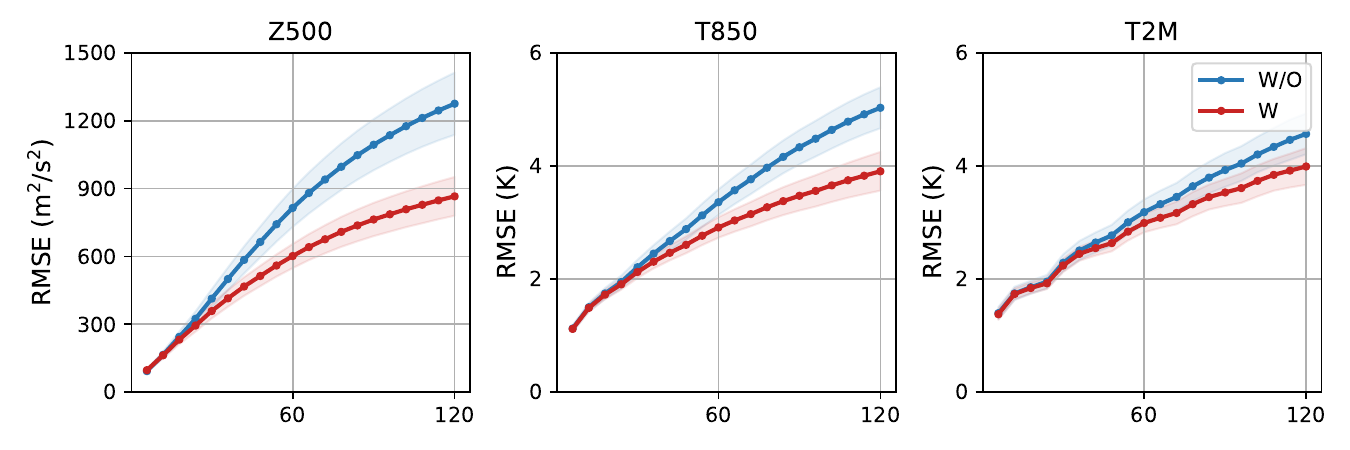}}
    \vskip -0.1in
    \caption{Validation of the iterative training strategy. W and W/O refer to with and without iterative training strategy.}
    \label{fig:iterative_training_strategy}
    \end{center}
    \vskip -0.2in
\end{figure}

To validate the improvement brought about by the iterative training process, we train the model for 150 epochs only to forecast 6 hours. The results shown in Figure \ref{fig:iterative_training_strategy} show the efficacy of the iterative training strategy. Although single-step training has comparable performance with iterative training on the first two time steps, and even surpasses the Z500 at a forecast time at 6 hours, it is less robust than iterative prediction in medium-range weather forecasting.

\subsection{Comparison of different graph edges}
\begin{figure}[h!]
    \begin{center}
    \centerline{\includegraphics[width=0.5\columnwidth]{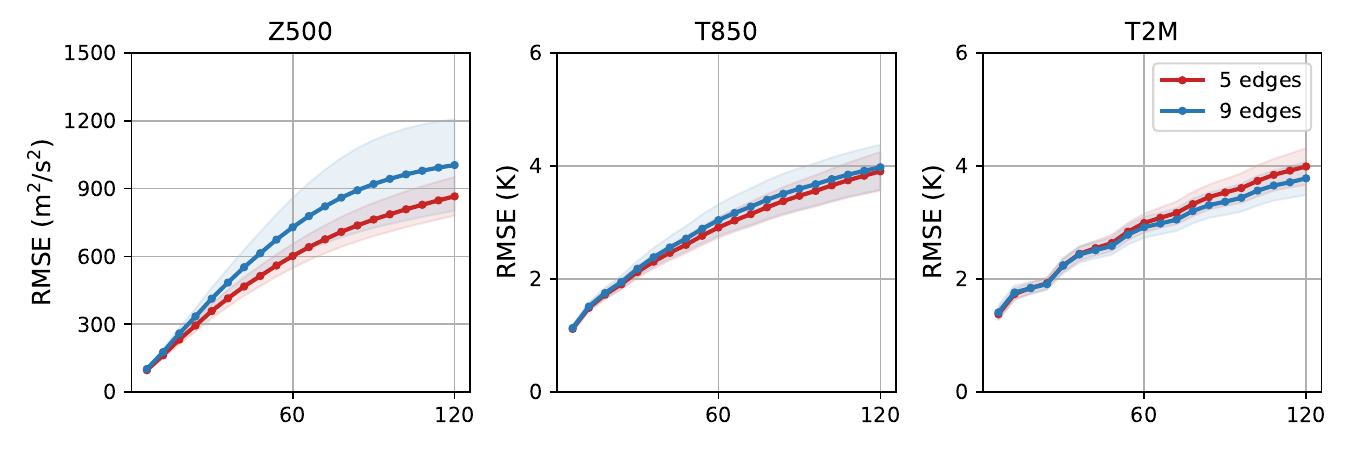}}
    \vskip -0.1in
    \caption{Comparison of graph with five edges and nine edges.}
    \label{fig:five_nine_comparison}
    \end{center}
    \vskip -0.2in
\end{figure}

We compare the performance of connecting the surrounding four nodes (five edges including a self-loop) and the surrounding eight nodes (nine edges) in our model, since, intuitively, connecting with all eight neighbors will incorporate more information in the dynamic processes, such as the wind blowing from northeast to southwest. The results are shown in Figure \ref{fig:five_nine_comparison}. Five-edge connection performs better than nine-edge connection in Z500 and T850, but not so well in T2M after five forecasting steps.

There are two reasons for us to adopt the five-edge structure. The first is that it demonstrates better performance in an overall situation. The second is from both data and atmospheric model levels. Since the horizontal winds in the atmospheric model, as well as the ERA5 dataset we used, are the meridional wind and zonal wind, we would like to follow the design of the Arakawa C-grid \cite{arakawa1977computational}, a widely used grid in modern atmospheric or weather forecasting models. In this case, the natural wind direction in each grid is calculated with meridional and zonal winds by the trigonometric function. Besides, the wind is the basic prognostic variable in the atmospheric model, and other factors, such as water vapor and aerosol, are transmitted by the wind. Therefore, using four nodes can describe the data given, while eight nodes will increase the uncertainty of variables when projected to the meridional and zonal direction.

\subsection{Analysis of different GAT layers}
\begin{figure}[h!]
    \begin{center}
    \centerline{\includegraphics[width=0.5\columnwidth]{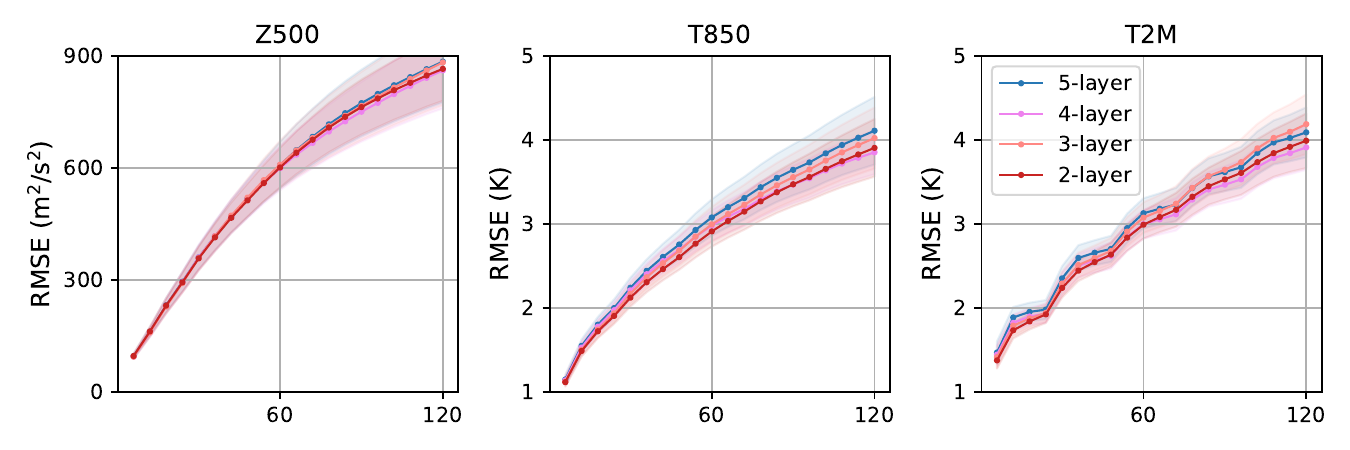}}
    \vskip -0.1in
    \caption{Analysis of the effects of different GAT layers.}
    \label{fig:diff_layers}
    \end{center}
    \vskip -0.2in
\end{figure}

GNNs have suffered from the over-smoothing problem, which means the node representations are indistinguishable during the message-passing process when multi-layers are stacked \cite{chen2020measuring}. Therefore, we test the performance of the proposed model with different graph attention layers ranging from 2 to 5, while other parts remain identical. The results are shown in Figure \ref{fig:diff_layers}.

The results demonstrate that while GAT with two layers performs best in the first 12 steps for Z500 and T2M, and the first 17 steps for T850, GAT with four layers is more stable in a long-term forecast. However, there is no significant difference in the performance of GAT across different layers.

\subsection{Competing Methods}
\begin{figure}[ht]
    \vskip 0.1in
    \begin{center}
    \centerline{\includegraphics[width=0.7\columnwidth]{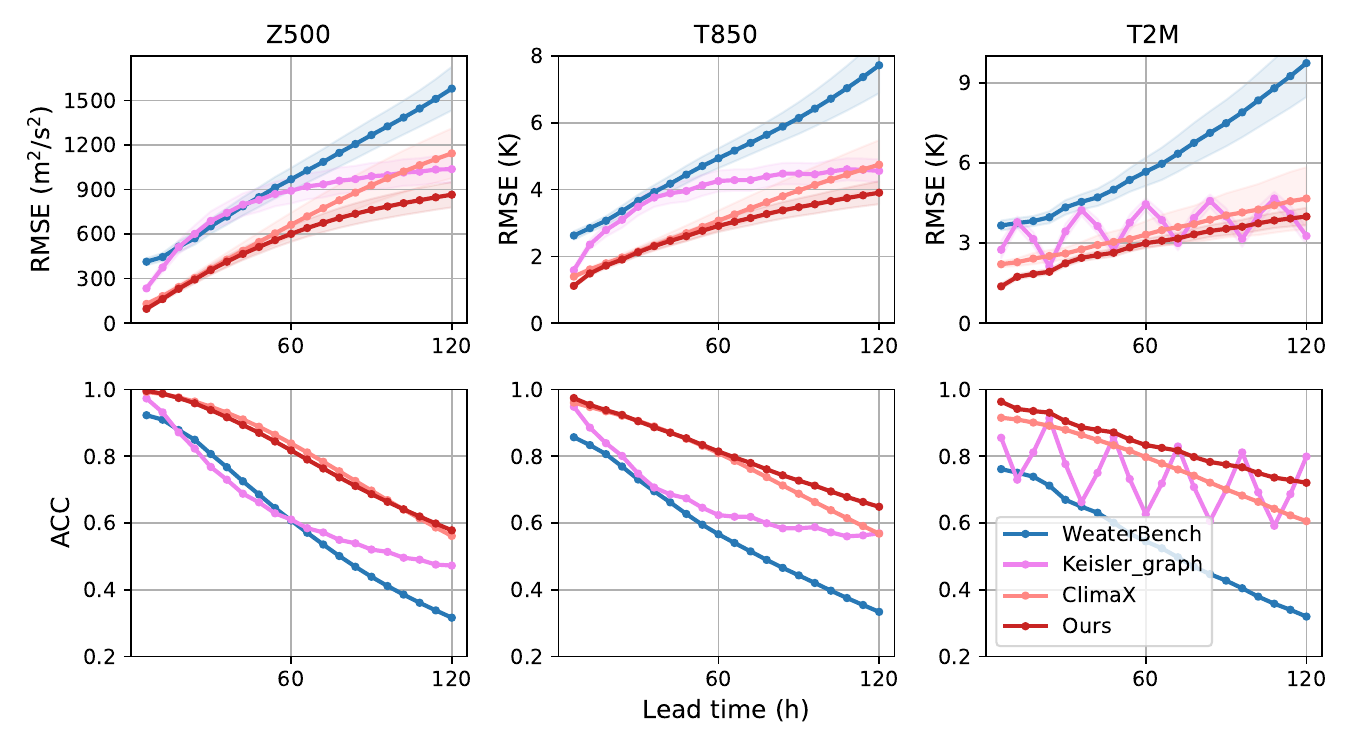}}
    \vskip -0.1in
    \caption{Comparison with competing methods for 5-day iterative forecasting.}
    \label{fig:competing}
    \end{center}
    \vskip -0.2in
\end{figure}

We conduct a comparative analysis of our model against several benchmark models, including a CNN-based model WeatherBench \cite{rasp2020weatherbench}, a graph-based model \cite{keisler2022forecasting}, and a Transformer-based model ClimaX \cite{nguyen2023climax}. These models vary in terms of the number of parameters and network scales. We train all these models from scratch using the same variables and training methods in this study to ensure the fairness and comparability of the results. One-step forecasting results are summarized in Table \ref{tab:competing}, and Figure \ref{fig:competing} compares the iterative forecasting performance. Our model outperforms these competing models according to these three variables, and is comparable to ClimaX with a much smaller number of parameters. We also include the NWP results from the Integrated Forecasting System \cite{ifswebsite} in Table \ref{tab:competing}. As expected, our method is not comparable to IFS at the current stage, and we will continuously refine our model in the future from both the model structure and the training method.

\begin{table}[h!]
  \begin{threeparttable}[b]
    \centering
    \caption{Comparison with competing methods for one-step prediction.}
    \vskip 0.1in
      \begin{tabular}{lrrrrrrrrc}
        \toprule
            & \multicolumn{2}{c}{Z500} & \multicolumn{2}{c}{T850} & \multicolumn{2}{c}{T2M} & \multicolumn{2}{c}{TP} & \multicolumn{1}{c}{\multirow{2}[0]{*}{Parameters (M)\tnote{1}}} \\
            \cline{2-9}
            & \multicolumn{1}{c}{RMSE$\downarrow$} & \multicolumn{1}{l}{ACC$\uparrow$} & \multicolumn{1}{l}{RMSE$\downarrow$} & \multicolumn{1}{l}{ACC$\uparrow$} & \multicolumn{1}{l}{RMSE$\downarrow$} & \multicolumn{1}{l}{ACC$\uparrow$} & \multicolumn{1}{l}{RMSE$\downarrow$} & \multicolumn{1}{l}{ACC$\uparrow$} &  \\
        \midrule
      \cite{rasp2020weatherbench}     & 413.98 & 0.93  & 2.62  & 0.88  & 3.66  & 0.82  & 0.42  & 0.39  &  6.37 \\
      \cite{keisler2022forecasting}     & 234.38    &   0.98 &  1.58 &  0.96  &  2.75 &  0.89  &  0.53  &  0.32  & \textcolor{Blue}{7.94} \\
      \cite{nguyen2023climax}     &  129.47 & 0.99  & 1.39  & 0.97  &  2.21 &  0.94 & 0.36  & 0.6 & \textcolor{Blue}{108.98} \\
      Ours     &   96.30 & 1.00 & 1.12 & 0.98 &  1.38 & 0.97  &  0.36  &  0.61  &  4.38 \\
      \midrule
      IFS     &   26.93 & 1.00 & 0.69 & 0.99 &  0.97 & 0.99  &  /  &  /  &  / \\
      \bottomrule
      \end{tabular}%
      \label{tab:competing}
      \begin{tablenotes}
        \item[1] Parameters in \textcolor{Blue}{blue} may be different from the original paper due to the adjustment for hyperparameters.
      \end{tablenotes}
    \end{threeparttable}%
    \vskip -0.2in
  \end{table}%
  
The discrepancy between our reproduced results and those in the original works can mainly be attributed to the different training methods, dataset composition, and resolution. For example, ClimaX initially incorporates an extensive volume of data from CMIP6 for pretraining, while we only use the ERA5 data to train it from scratch. Besides, the difference in training datasets introduces variations in learned representations. \cite{keisler2022forecasting} uses single timestep variables (78 variables with 1$^{\circ}$ resolution) as the input, while we used variables from the previous three timesteps (137 variables with 5.625$^{\circ}$ resolution). The decreased resolution may result in sparse information when mapped to the icosahedron mesh with a total number of 5,882 cells, and the information becomes over-smoothing after nine rounds of message-passing GNN.

To sum up, according to the results of both the ablation study and the comparison with competing methods, our proposed GAT-MLP model can well capture the global atmospheric patterns, and reflect the variation of temperature diurnal cycle with a much smaller number of parameters. It also shows the great potential of using the combination of graph and MLP to simulate the dynamics and physics process in weather forecasting by learning the intricate interactions between the variables and locations.

\section{Discussion}
In this work, we proposed a simple yet effective two-stage GAT-MLP network to simulate the weather forecasting process. The first component is a 2-layer GAT to simulate the advection between the interconnected neighbors. The second component is a 4-layer MLP to learn the convection within each vertical column. Previous studies prefer to use a unified model to substitute the whole weather forecasting process, and most of them discarding the original development of numerical models, while we split the network into two parts after rethinking the effectiveness of integrating the dynamics and physics in the atmospheric model. This work demonstrates the potential of using a lightweight model to represent the complex process in atmosphere by learning from numerical models. However, there are still some limitations that require further investigation.

\paragraph{Resolution} The resolution of the data is 5.625$^{\circ}$, which is not comparable to the results from the operational weather forecasting models such as ECMWF's IFS and NOAA's GFS in terms of both resolution and accuracy. Although 5.625$^{\circ}$ resolution is able to describe the general global weather patterns with minimal information loss \cite{rasp2020weatherbench}, and can validate the effectiveness of our proposed method, it may fall short of capturing the small-scale weather phenomena. Several studies point out that the model performance will be enhanced with the increase in data resolution \cite{rasp2021data,nguyen2023climax}. However, it is also challenging to push the resolution to a higher level due to the computational costs and memory limitations.

\paragraph{Physical Meanings} 
Currently, all these variables are put into the model at the very beginning without considering the physical meanings, the explicit vertical dependence in stencil calculation, and the sequential order of variable calculation in a real atmospheric model. In the future, we would like to split the variables according to their association with the dynamic core and physical processes, such as the recently published work \cite{kochkov2023neural}. For example, variables like wind components predominantly rely on the dynamic core for calculation, while precipitation and radiation are calculated in the physical processes. Also, physical meanings can be added to the GAT layers to ensure the conservation. The future direction aligns with our commitment to refine the model's representation of the atmospheric model and enhance its consistency with the physical constraints.

In summary, this work explores the potential of decomposing the weather forecasting model into advection and convection. Specifically, the flexibility of graph structures has provided our model with significant room for development, which we will explore in the future.


\newpage
\appendix
\section{Data}
We downloaded the preprocessed ERA5 data from \url{https://dataserv.ub.tum.de/index.php/s/m1524895} \cite{rasp2020weatherbench}. The 5.625$^{\circ}$ resolution data is regridded from 0.25$^{\circ}$ resolution. We use 5.625$^{\circ}$ resolution dataset for all the experiments in this work.

\section{Grid to Node Conversion}\label{grid2node_appendix}
\begin{figure*}[h!]
    \vskip 0.2in
    \begin{center}
    \centerline{\includegraphics[width=0.5\columnwidth]{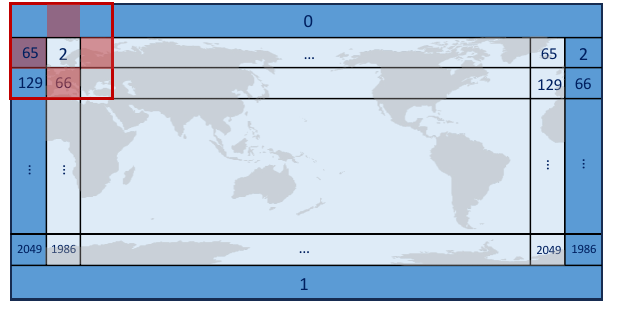}}
    \caption{The grid to node conversion. Areas in light blue is the original $32 \times 64$ grid, and areas in dark blue is the padded grid.}
    \label{fig:grid2node}
    \end{center}
    \vskip -0.2in
\end{figure*}

The process of converting from grids to nodes is shown in Figure \ref{fig:grid2node}. The number in each grid represents the node index in the graph. We pad the original area with $32\times 64$ grids to $34\times 66$ grids, and connect each gird in the original area to its four neighbours (top, bottom, left and right grids). Therefore, the number of nodes in the graph is $32\times 64 + 2 = 2050$ in total.

\section{Codes for Competing Methods}
In this study, we use the WeatherBench \cite{rasp2020weatherbench}, a graph-based model \cite{keisler2022forecasting}, and ClimaX \cite{nguyen2023climax}. The codes are adapted from the source code in GitHub repositories listed in Table \ref{tab:competing_codes}.
The variables used, the training and validation process are identical to our proposed model, after adjusting the parameters and initial learning rate for each model.

\begin{table*}[h!]
    \centering
    \caption{Source code links for competing methods}
    \vskip 0.1in
      \begin{tabular}{ll}
        \toprule
        Methods & Links \\
        \midrule
            \cite{rasp2020weatherbench} & \url{https://github.com/raspstephan/WeatherBench.git} \\
            \cite{keisler2022forecasting} & \url{https://github.com/openclimatefix/graph_weather.git} \\
            \cite{nguyen2023climax} & \url{https://github.com/microsoft/ClimaX.git} \\
      \bottomrule
      \end{tabular}%
    \label{tab:competing_codes}%
    \vskip -0.1in
  \end{table*}%

\subsection{Hyperparameters}
Table \ref{tab:hyper_weatherbench} to Table \ref{tab:hyper_climax} listed the hyperparameters we use in the competing methods. We mainly adopt the default parameters from the source code, and modify the start learning rate to a suitable value.


\begin{table}[h!]
  \centering
  \caption{Hyperparameters of WeatherBench \cite{rasp2020weatherbench}}
  \vskip 0.1in
    \begin{tabular}{ll}
      \toprule
    Hyperparameter & \multicolumn{1}{l}{Value} \\
    \midrule
    Image size & \multicolumn{1}{l}{[32,64]} \\
    Padding & \multicolumn{1}{l}{PeriodicPadding} \\
    Kernel size & 3 \\
    Stride & 1 \\
    Hidden dimension & 128 \\
    Residual blocks & 19 \\
    Dropout & 0.1 \\
    Activation & LeakyReLU \\
    \bottomrule
    \end{tabular}%
  \label{tab:hyper_weatherbench}%
  \vskip -0.1in
\end{table}%

\begin{table}[h!]
  \centering
  \caption{Hyperparameters of GraphWeather \cite{keisler2022forecasting}}
  \vskip 0.1in
    \begin{tabular}{ll}
      \toprule
    Hyperparameter & \multicolumn{1}{l}{Value} \\
    \midrule
    Resolution for h3 mapping & 2 \\
    Node dimension & 256 \\
    Edge dimension & 256 \\
    Number of blocks in message passing & 9 \\
    Hidden dim in processor node & 256 \\
    Hidden dim in process edge & 256 \\
    Hidden layers in processor node & 2 \\
    Hidden layers in processor edge & 2 \\
    Hidden dim in encoder & 128 \\
    Hidden layers in decoder & 2 \\
    \bottomrule
    \end{tabular}%
  \label{tab:hyper_graphweather}%
  \vskip -0.1in
\end{table}%

\begin{table}[h!]
  \centering
  \caption{Hyperparameters of ClimaX \cite{nguyen2023climax}}
  \vskip 0.1in
    \begin{tabular}{ll}
      \toprule
    Hyperparameter & \multicolumn{1}{l}{Value} \\
    \midrule
    Image size & \multicolumn{1}{l}{[32,64]} \\
    Patch size & 2 \\
    Embedding dimension & 1024 \\
    Number of ViT blocks & 8 \\
    Number of attention heads & 16 \\
    MLP ratio & 4 \\
    Prediction depth & 2 \\
    Hidden dimension in prediction head & 1024 \\
    Drop path & 0.1 \\
    Drop rate & 0.1 \\
    \bottomrule
    \end{tabular}%
  \label{tab:hyper_climax}%
  \vskip -0.1in
\end{table}%

\section{Codes for Calculating Parameters}
We use the following codes to calculate the number of parameters for each model.
\begin{lstlisting}[language=Python]
  def count_parameters(self):
    return sum(para.numel() for para in self.model.parameters() if para.requires_grad)
\end{lstlisting}

\section{Software and Hardware}
The model is built with PyTorch \cite{paszke2019pytorch} with the version of PyTorch1.13.1+cu117. The Graph Attention Network is built with PyTorch Geometric 2.4.0. All the experiments are conducted on NVIDIA RTX 4090 GPUs with 24GB memory.

\end{document}